\documentclass[twocolumn,prl,unsortedaddress,superscriptaddress,letter]{revtex4}

\usepackage{graphicx}
\usepackage{dcolumn}
\usepackage{amsmath}

\begin{document}


\author{R.U.Abbasi}
\affiliation{University of Utah,
Department of Physics and High Energy Astrophysics Institute,
Salt Lake City, Utah, USA}

\author{T.~Abu-Zayyad}
\affiliation{University of Utah,
Department of Physics and High Energy Astrophysics Institute,
Salt Lake City, Utah, USA}

\author{J.F.~Amann}
\affiliation{Los Alamos National Laboratory,
Los Alamos, NM, USA}

\author{G.~Archbold}
\affiliation{University of Utah,
Department of Physics and High Energy Astrophysics Institute,
Salt Lake City, Utah, USA}

\author{J.A.~Bellido}
\affiliation{University of Adelaide, Department of Physics,
Adelaide, South Australia,  Australia}

\author{K.~Belov}
\affiliation{University of Utah,
Department of Physics and High Energy Astrophysics Institute,
Salt Lake City, Utah, USA}

\author{J.W.~Belz}
\affiliation{University of Montana, Department of Physics and Astronomy,
Missoula, Montana, USA.}

\author{D.R.~Bergman}
\affiliation{Rutgers --- The State University of New Jersey,
Department of Physics and Astronomy,
Piscataway, New Jersey, USA}

\author{Z.~Cao}
\affiliation{University of Utah,
Department of Physics and High Energy Astrophysics Institute,
Salt Lake City, Utah, USA}

\author{R.W.~Clay}
\affiliation{University of Adelaide, Department of Physics,
Adelaide, South Australia,  Australia}

\author{M.D.~Cooper}
\affiliation{Los Alamos National Laboratory,
Los Alamos, NM, USA}

\author{H.~Dai}
\affiliation{University of Utah,
Department of Physics and High Energy Astrophysics Institute,
Salt Lake City, Utah, USA}

\author{B.R.~Dawson}
\affiliation{University of Adelaide, Department of Physics,
Adelaide, South Australia,  Australia}

\author{A.A.~Everett}
\affiliation{University of Utah,
Department of Physics and High Energy Astrophysics Institute,
Salt Lake City, Utah, USA}

\author{J.H.V.~Girard}
\affiliation{University of Utah,
Department of Physics and High Energy Astrophysics Institute,
Salt Lake City, Utah, USA}

\author{R.C.~Gray}
\affiliation{University of Utah,
Department of Physics and High Energy Astrophysics Institute,
Salt Lake City, Utah, USA}

\author{W.F.~Hanlon}
\affiliation{University of Utah,
Department of Physics and High Energy Astrophysics Institute,
Salt Lake City, Utah, USA}

\author{C.M.~Hoffman}
\affiliation{Los Alamos National Laboratory,
Los Alamos, NM, USA}

\author{M.H.~Holzscheiter}
\affiliation{Los Alamos National Laboratory,
Los Alamos, NM, USA}

\author{P.~H\"{u}ntemeyer}
\affiliation{University of Utah,
Department of Physics and High Energy Astrophysics Institute,
Salt Lake City, Utah, USA}

\author{B.F Jones}
\affiliation{University of Utah,
Department of Physics and High Energy Astrophysics Institute,
Salt Lake City, Utah, USA}

\author{C.C.H.~Jui}
\affiliation{University of Utah,
Department of Physics and High Energy Astrophysics Institute,
Salt Lake City, Utah, USA}

\author{D.B.~Kieda}
\affiliation{University of Utah,
Department of Physics and High Energy Astrophysics Institute,
Salt Lake City, Utah, USA}

\author{K.~Kim}
\affiliation{University of Utah,
Department of Physics and High Energy Astrophysics Institute,
Salt Lake City, Utah, USA}

\author{M.A.~Kirn}
\affiliation{University of Montana, Department of Physics and Astronomy,
Missoula, Montana, USA.}

\author{E.C.~Loh}
\affiliation{University of Utah,
Department of Physics and High Energy Astrophysics Institute,
Salt Lake City, Utah, USA}

\author{N.~Manago}
\affiliation{University of Tokyo,
Institute for Cosmic Ray Research,
Kashiwa, Japan}

\author{L.J.~Marek}
\affiliation{Los Alamos National Laboratory,
Los Alamos, NM, USA}

\author{K.~Martens}
\affiliation{University of Utah,
Department of Physics and High Energy Astrophysics Institute,
Salt Lake City, Utah, USA}

\author{G.~Martin}
\affiliation{University of New Mexico,
Department of Physics and Astronomy,
Albuquerque, New Mexico, USA  }

\author{J.A.J.~Matthews}
\affiliation{University of New Mexico,
Department of Physics and Astronomy,
Albuquerque, New Mexico, USA  }

\author{J.N.~Matthews}
\affiliation{University of Utah,
Department of Physics and High Energy Astrophysics Institute,
Salt Lake City, Utah, USA}

\author{S.~Meltzer}
\affiliation{Columbia University, Department of Physics and
Nevis Laboratory, New York, New York, USA}

\author{J.R.~Meyer}
\affiliation{University of Utah,
Department of Physics and High Energy Astrophysics Institute,
Salt Lake City, Utah, USA}

\author{S.A.~Moore}
\affiliation{University of Utah,
Department of Physics and High Energy Astrophysics Institute,
Salt Lake City, Utah, USA}

\author{P.~Morrison}
\affiliation{University of Utah,
Department of Physics and High Energy Astrophysics Institute,
Salt Lake City, Utah, USA}

\author{A.N.~Moosman}
\affiliation{University of Utah,
Department of Physics and High Energy Astrophysics Institute,
Salt Lake City, Utah, USA}

\author{J.R.~Mumford}
\affiliation{University of Utah,
Department of Physics and High Energy Astrophysics Institute,
Salt Lake City, Utah, USA}

\author{M.W.~Munro}
\affiliation{University of Montana, Department of Physics and Astronomy,
Missoula, Montana, USA.}

\author{C.A.~Painter}
\affiliation{Los Alamos National Laboratory,
Los Alamos, NM, USA}

\author{L.~Perera}
\affiliation{Rutgers --- The State University of New Jersey,
Department of Physics and Astronomy,
Piscataway, New Jersey, USA}

\author{K.~Reil}
\affiliation{University of Utah,
Department of Physics and High Energy Astrophysics Institute,
Salt Lake City, Utah, USA}

\author{R.~Riehle}
\affiliation{University of Utah,
Department of Physics and High Energy Astrophysics Institute,
Salt Lake City, Utah, USA}

\author{M.~Roberts}
\affiliation{University of New Mexico,
Department of Physics and Astronomy,
Albuquerque, New Mexico, USA  }

\author{J.S.~Sarracino}
\affiliation{Los Alamos National Laboratory,
Los Alamos, NM, USA}

\author{M.~Sasaki}
\affiliation{University of Tokyo,
Institute for Cosmic Ray Research,
Kashiwa, Japan}

\author{S.R.~Schnetzer}
\affiliation{Rutgers --- The State University of New Jersey,
Department of Physics and Astronomy,
Piscataway, New Jersey, USA}

\author{P.~Shen}
\affiliation{University of Utah,
Department of Physics and High Energy Astrophysics Institute,
Salt Lake City, Utah, USA}

\author{K.M.~Simpson}
\affiliation{University of Adelaide, Department of Physics,
Adelaide, South Australia,  Australia}

\author{G.~Sinnis}
\affiliation{Los Alamos National Laboratory,
Los Alamos, NM, USA}

\author{J.D.~Smith}
\affiliation{University of Utah,
Department of Physics and High Energy Astrophysics Institute,
Salt Lake City, Utah, USA}

\author{P.~Sokolsky}
\affiliation{University of Utah,
Department of Physics and High Energy Astrophysics Institute,
Salt Lake City, Utah, USA}

\author{C.~Song}
\affiliation{Columbia University, Department of Physics and
Nevis Laboratory, New York, New York, USA}

\author{R.W.~Springer}
\affiliation{University of Utah,
Department of Physics and High Energy Astrophysics Institute,
Salt Lake City, Utah, USA}

\author{B.T.~Stokes}
\affiliation{University of Utah,
Department of Physics and High Energy Astrophysics Institute,
Salt Lake City, Utah, USA}

\author{S.F.~Taylor}
\affiliation{University of Utah,
Department of Physics and High Energy Astrophysics Institute,
Salt Lake City, Utah, USA}

\author{S.B.~Thomas}
\affiliation{University of Utah,
Department of Physics and High Energy Astrophysics Institute,
Salt Lake City, Utah, USA}

\author{T.N.~Thompson}
\affiliation{Los Alamos National Laboratory,
Los Alamos, NM, USA}

\author{G.B.~Thomson}
\affiliation{Rutgers --- The State University of New Jersey,
Department of Physics and Astronomy,
Piscataway, New Jersey, USA}

\author{D.~Tupa}
\affiliation{Los Alamos National Laboratory,
Los Alamos, NM, USA}

\author{S.~Westerhoff}
\affiliation{Columbia University, Department of Physics and
Nevis Laboratory, New York, New York, USA}

\author{L.R.~Wiencke}
\affiliation{University of Utah,
Department of Physics and High Energy Astrophysics Institute,
Salt Lake City, Utah, USA}

\author{T.D.~VanderVeen}
\affiliation{University of Utah,
Department of Physics and High Energy Astrophysics Institute,
Salt Lake City, Utah, USA}

\author{A.~Zech}
\affiliation{Rutgers --- The State University of New Jersey,
Department of Physics and Astronomy,
Piscataway, New Jersey, USA}

\author{X.~Zhang}
\affiliation{Columbia University, Department of Physics and
Nevis Laboratory, New York, New York, USA}

\collaboration{The High Resolution Fly's Eye Collaboration}

\title{ Measurement of the Flux of Ultrahigh Energy
  Cosmic Rays from Monocular Observations by the High
  Resolution Fly's Eye Experiment.}

\begin{abstract}
We have measured the cosmic ray spectrum above $10^{17.2}$~eV using the
two air fluorescence detectors of the High Resolution Fly's Eye
observatory operating in monocular mode.  We describe the detector,
photo-tube and atmospheric calibrations, as well as the analysis
techniques for the two detectors.  We fit the spectrum to a model
consisting of galactic and extra-galactic sources.  
\end{abstract}

\maketitle

The highest energy cosmic rays detected so far, of energies up to and
exceeding $10^{20}$~eV, are very interesting in that they shed light on two
important questions: the nature of their origin in astrophysical or other
sources and their propagation to us through the Cosmic Microwave Background
Radiation (CMBR). The production of pions from interactions of CMBR photons
and Ultra High Energy Cosmic Rays (UHECR) is an important energy loss
mechanism above $\sim10^{19.8}$ eV, and produces the
Greisen-Zatsepin-K'uzmin (GZK) effect
\cite{gr,zk}; $e^+ e^-$ production in the same collisions is a
weaker energy-loss mechanism above a threshold of $10^{17.8}$~eV. We report
here the flux of UHECR from $10^{17.2}$ eV to over
$10^{20}$ eV, measured in monocular mode, with the High Resolution Fly's Eye
(HiRes) detectors.

The HiRes observatory consists of two air-fluorescence detector sites
separated by 12.6~km and located at the U.S. Army Dugway Proving Ground in
Utah. Cosmic rays interacting in the upper atmosphere initiate particle
cascades known as extensive air-showers. Passage of charged particles
excites nitrogen molecules causing emission of (mostly) ultraviolet light. 
The fluorescence yield has been previously measured by Kakimoto {\it et
al.}~\cite{kakimoto}, and more recently by Nagano {\it et al.}~\cite{nagano}
For this analysis, we used the fluorescence spectrum compiled by
Bunner~\cite{bunner} and normalized it to the yield of Kakimoto.  By
measuring the longitudinal development of the fluorescence signal, one can
infer the arrival direction, energy and average composition of the primary
cosmic ray.  HiRes was designed to measure the fluorescence light
stereoscopically. However our two detectors trigger and reconstruct events
independently. In this ``monocular'' mode our current data have
significantly better statistical power and cover a much wider energy range
than our stereo sample.

The two HiRes detector sites, referred to as HiRes-I and HiRes-II, are
operated on clear, moon-less nights. Over a typical year, each detector
accumulates up to 1000 hours of observation. The HiRes-I site has been in
operation since June of 1997~\cite{hr1_det}. It consists of 21 detector
units, each equipped with a $5~$m$^{2}$ spherical mirror and 256 photo-tube
pixels at its focal plane. Each photo-tube covers a $1^{\circ}$ cone of sky.
These 21 mirrors cover elevation angles between 3$^{\circ}$ and
17$^{\circ}$.  The HiRes-I electronics perform sample-and-hold integration
in a 5.6~$\mu$s window, which is long enough to contain signals from all
reconstructible events. The HiRes-II site was completed in late 1999 and
began observations that year. This site uses the same type of mirrors and
photo-tubes as HiRes-I, but contains 42 mirrors, in two rings, covering
elevation angles from 3$^{\circ}$ to 31$^{\circ}$.  HiRes-II uses an FADC
data acquisition system operating at 10~MHz~\cite{hr2_det}. Both the HiRes-I
and -II sites provide $2\pi$ azimuthal angle coverage.

To determine the correct shower energies, the air fluorescence
technique requires accurate measurement and monitoring of photo-tube
gains. Two methods of calibration are used. Pulses from a YAG laser
are distributed to mirrors via optical fibers. They provide a nightly
relative calibration.  A stable, standard light source is used for a
more precise monthly absolute calibration. Overall, the relative
photo-tube gains were stable to within 3.5\% and the absolute gains
were known to $\pm{10}\%$~\cite{hr_cal}.

A second variable in the energy measurement is atmospheric clarity. Light
from air showers is attenuated by: (a)~molecular (Rayleigh), and (b)~aerosol
scattering. The former is approximately constant, subject only to small
variations in the atmospheric overburden.  The aerosol concentration varies
with time. HiRes measures the aerosol content by observing scattered light
from two steerable laser systems. The laser observed by HiRes-I has been in
operation since 1999. The vertical aerosol optical depth from ground to
3.5~km altitude, $\tau_{A}$, is measured each hour (the vertical
transmission through the aerosol is $T=e^{-\tau_{A}}$). Over two years,
these measurements yielded an average $\tau_{A}$ at 355~nm of
0.04~\cite{hr_atm}.  The RMS of the distribution is 0.02, and the systematic
uncertainty in the mean is no larger than this. The average aerosol
ground-level horizontal extinction length, $\Lambda_{H}$, was determined to
be 25~km.  Because about half of our data were collected before the
steerable lasers were in operation we used these averages in our analysis
and simulation.

Between June 1997 and February 2003, the HiRes-I detector operated for
approximately 3600 hours. From this, 2820 hours of good weather data were
analyzed. We selected 5.5 million downward, track-like events. For each of
these, a shower-detector plane was determined from the pattern of photo-tube
hits. We excluded events containing an average number of photo-electrons per
photo-tube of less than 25, where fluctuations in signals are too great to
permit reliable reconstruction. We also cut out tracks with angular speed in
excess of $3.33^{\circ}/\mu$s; for these events (typically within 5~km of
HiRes-I) the shower maxima appear above the field of view. We selected
12,709 events for reconstruction.

Determination of the shower geometry is possible in monocular mode. 
The impact parameter, $R_p$, and the angle of
the shower in the plane containing the shower and the detector, $\psi$, are
found by fitting the photo-tube trigger times to the angles at which
they view the shower.  However, HiRes-I monocular events are too short in
angular spread for reliable pure-timing fit. For this analysis, the expected
form of the shower development was used to constrain the time fit to
yield realistic geometries. The shower profile is assumed to be described by
the Gaisser-Hillas parameterization~\cite{gh}, which has been found to be in
good agreement with previous HiRes measurements~\cite{hr_prof} and with
CORSIKA/QGSJET simulations~\cite{csong,corsika,qgsjet}. This technique is
called the Profile-Constrained Fit (PCF). We allowed the shower maximum,
$x_{m}$ to vary in 35~g/cm$^{2}$ steps between 680 and 900~g/cm$^{2}$,
matching the expected range of $x_{m}$ for proton to iron primaries.
After reconstruction, we require a minimum track arc length of
8.0$^{\circ}$ and a maximum depth for the highest elevation hit of
1000~g/cm$^{2}$.  Significant contamination from the forward-beamed direct
\v{C}erenkov light degrades the reliability of the PCF. Therefore, we
rejected tracks with $\psi>120^{\circ}$ and those with two or more angular
bins of the shower with $>$25\% \v{C}erenkov light.
A total of 6,920 events remained.

Monte Carlo studies were performed to assess the reliability of the PCF
method. The simulated events were subjected to the same selection criteria
and cuts imposed on the data. An RMS energy resolution of better than 20\%
was seen above $10^{19.5}$~eV. However, the resolution degrades at
lower energies to about 25\% at $10^{18.5}$~eV. These Monte Carlo
results were cross-checked by examination of a smaller set of stereo events
where the geometry is more precisely known. Comparing the energies
reconstructed using monocular and stereo geometries, we obtained resolutions
similar to those seen in simulation.

\begin{figure}
\includegraphics[width=\columnwidth]{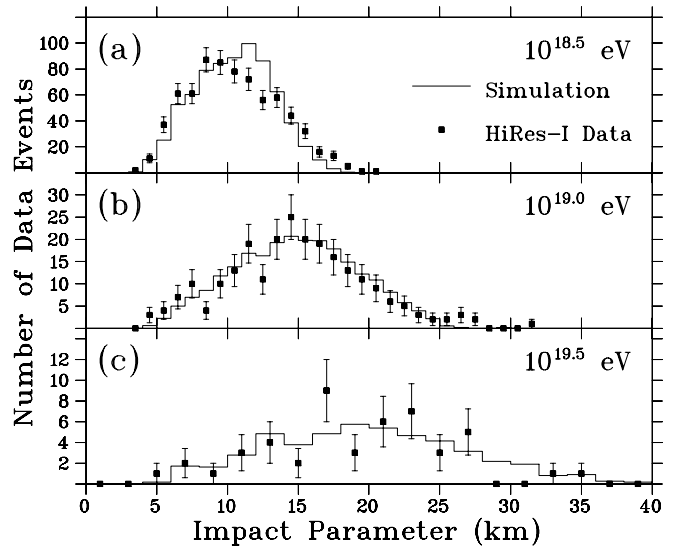}
  \caption{Comparison of HiRes-I simulated (histogram) and
           observed (points) $R_{p}$ distributions at (a) $10^{18.5}$, (b)
           $10^{19.0}$, and (c) $10^{19.5}$ eV. The Monte Carlo distributions
           are normalized to the number of data events.}
  \label{rp_comp}
\end{figure}

The simulation is also used to calculate the aperture.  To verify the
reliability of this calculation, we compared Monte Carlo distributions
of many geometrical and physical variables to the actual data, and
consistently found good agreement.  The Monte Carlo predictions for
the zenith angle and impact parameter ($R_p$), in particular, are
sensitive to the detector operating parameters.  We use input
parameters representative of actual running conditions, and again see
good agreement between data and simulation.  For example, we show the
comparison of $R_p$ distribution at three energies in Figure~\ref{rp_comp}.

The analysis of HiRes-II monocular data was similar to that for
HiRes-I.  The data sample was collected during 142 hours of good
weather between Dec.  1999 and May 2000. This period represents the
first stable running of the HiRes-II detector. At the end of this
period, a considerable change was made in the trigger, so that
subsequently collected data will be analyzed separately. With the
greater elevation coverage at HiRes-II, it was feasible to reconstruct
the shower geometry from timing alone (the PCF is unnecessary).
Therefore we were able to loosen some cuts for the HiRes-II fits. At
this stage 104,048 downward-going events remained.

\begin{figure}[floatfix]
  \includegraphics[width=\columnwidth]{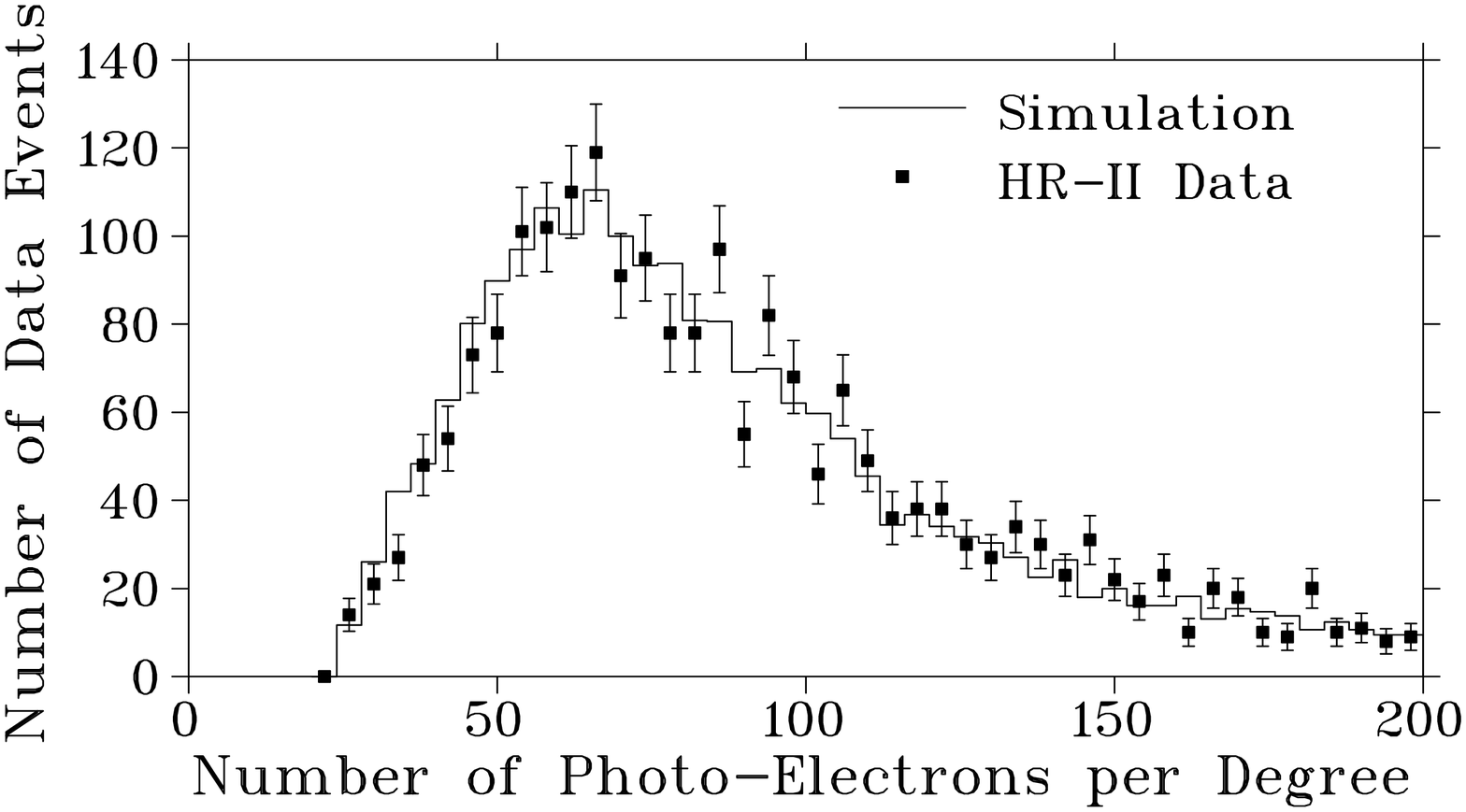}
  \caption{A comparison of the number of photo-electrons per degree of track
seen in HiRes-II monocular events (data points) and in simulation
(histogram).
 }
  \label{npedeg}
\end{figure}

With the geometry of the shower known, we fit the measured shower profile
to the Gaisser-Hillas parameterization~\cite{gh}. The events were required
to have a good fit to the Gaisser-Hillas function, to have a track length
greater than 10$^\circ$ for upper ring or multi-mirror events, a track
length greater than 7$^\circ$ for lower ring events, an angular speed less
than $11^{\circ}/\mu$s (the larger cut for HiRes-II reflects its extended
elevation coverage), a zenith angle less than 60$^\circ$, and a shower
maximum visible in our detector.  There was also a cut on the size of the
\v{C}erenkov light subtraction at $<$ 60\% of signal. Again, the same
selections and cuts were applied to both simulated and real events.  There
were 781 events left after these cuts. These simulations also gave excellent
reproduction of the data, as seen, for example, in the comparison of the
number of photo-electrons per degree of track in Fig.~\ref{npedeg}.

For both HiRes-I and HiRes-II events, the photo-electron count was converted
to a shower size at each atmospheric depth, using the known geometry of
the shower, and corrected for atmospheric attenuation. We integrated the
resulting function over $x$
and then multiplied by the average energy loss per
particle to give the visible shower energy.
A correction for energy carried off by non-observable
particles to give the total shower energy ($\sim10\%$)~\cite{csong}
was then applied.

\begin{figure}[floatfix]
  \includegraphics[width=\columnwidth]{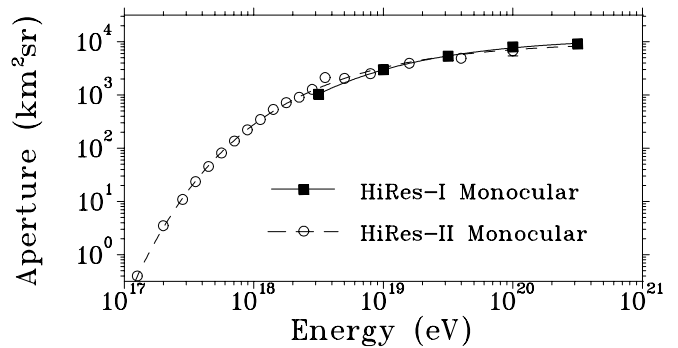}
  \caption{
    Calculated HiRes monocular Reconstruction aperture in the energy
    range ${10}^{17}-{10}^{20.5}$~eV. The HiRes-I and II apertures
    are shown by the squares and circles, respectively.}
  \label{aper}
\end{figure}

The monocular reconstruction apertures are shown in
Fig.~\ref{aper}. Both approach $10^{4}$~km$^{2}$sr above
$10^{20}$~eV. We restrict our result for HiRes-I to energies
$>{10}^{18.5}$~eV; below this the PCF technique is
unstable. Due to longer tracks and additional timing information, the
RMS energy resolution for HiRes-II remains better than 30\% down to
$10^{17.2}$~eV. However, the HiRes-II data become statistically
depleted above $10^{19}$~eV.

\begin{figure}
  \includegraphics[width=\columnwidth]{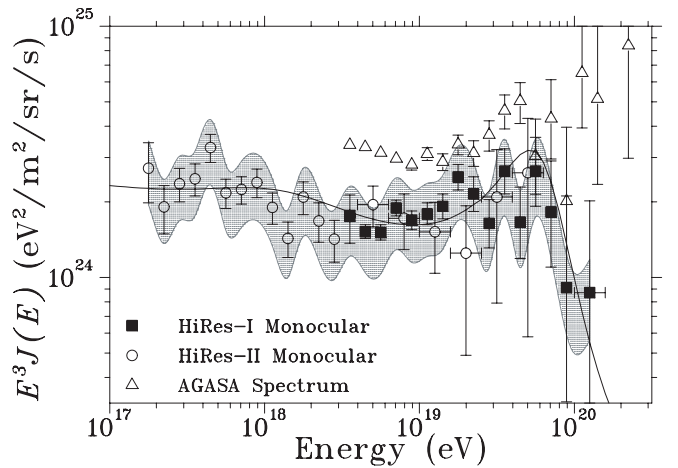}
  \caption{
    Combined HiRes monocular spectrum. The squares and circles
    represent the HiRes-I and II differential flux $J(E)$, multiplied by
    $E^{3}$. The error bars are statistical only, and the systematic
    uncertainties are indicated by the shaded region. The line is a
    fit to the data of a model, described in the text, of galactic and
    extra-galactic cosmic ray sources. The AGASA spectrum~\cite{agasa}
    is shown by triangles for comparison.}
  \label{spec1}
\end{figure}

We calculated the cosmic ray flux for HiRes-I above
${10}^{18.5}$~eV, and for HiRes-II above
${10}^{17.2}$~eV. This combined spectrum is shown in
Fig.~\ref{spec1}, where the flux $J(E)$ has been multiplied by
$E^{3}$. The error bars represent
the 68\% confidence interval for the Poisson fluctuations in the
number of events. The HiRes-I flux is the result of two
independent analyses~\cite{tareq,zhang}.

The largest systematic uncertainties in the energy scale are the absolute
calibration of the photo-tubes ($\pm{10}$\%)~\cite{hr_cal}, the fluorescence
yield ($\pm{10}$\%)~\cite{kakimoto,nagano}, and the correction for
unobserved energy in the shower ($\pm5$\%)~\cite{csong,linsley}. Excluding
atmospheric effects, the energy scale uncertainty is
$\pm{15}$\%. This translates to a systematic uncertainty in the flux,
$J(E)$, of $\pm{27}$\%.

We estimate the atmospheric contribution to the energy error by repeating
the event reconstruction with $\tau_{A}$ varied by $\pm{1}$ RMS value, from
0.04 to 0.06 and 0.02. 
While a $\pm{0.02}$ change in $\tau_{A}$ represents a $\pm{50}$\% change in
aerosol concentration, that contribution to the attenuation at these levels
is small, and a 0.02 change modifies the transmission by only $\sim{10}$\%
at 25 km from the detector. We found the
reconstructed geometries of HiRes-I events above
$10^{18.5}$~eV to be insensitive to changes in assumed $\tau_{A}$, and we
saw a maximum change in the energy of $\pm{13}$\% at
$10^{20}$~eV, decreasing to $\pm{6}$\% at $10^{18.5}$~eV. The geometries of the
HiRes-II events do not depend on $\tau_{A}$ at all, and the change in energy
scale for these are typically 6\% or less. Taking the HiRes-I average energy
shift, 9\%, the overall systematic uncertainty in energy scale, including
atmospheric effects, then becomes $\pm{17}$\%.

We also recalculated the aperture and the flux with
$\tau_{A}$ changed by $\pm{0.02}$. From these we obtained an average
atmospheric uncertainty in $J(E)$ of $\pm{15}$\%, and an uncertainty in the
flux of $\pm{31}$\%. The overall systematic uncertainties in the flux,
including the energy-dependent atmospheric contribution (with a slope in
$|\Delta{J}(E)|$ of about 5\% per decade of energy) are indicated by the
shaded region in Figure~\ref{spec1}. The relative calibration uncertainty
between the two detector sites is less than 10\%.

Our data contain two events at or above ${10}^{20}$~eV, measured at $1.0$
and $1.5\times{10}^{20}$~eV. Assuming a purely molecular atmosphere
($\tau_{A}=0.0$), we obtain lower energy limits of
$0.9$ and $1.2\times{10}^{20}$~eV. In the energy range where both
detectors' data have good statistical power, the results agree with each
other very well. However, our flux values are on average 13\% lower than the
stereo spectrum reported by Fly's Eye~\cite{fester}. This
difference can be explained by a 7\% offset in the energy calibration alone,
well within the stated uncertainty of the two experiments. 

The GZK effect predicts a suppression in the UHECR flux above
$10^{19.8}$~eV. We fit our data to a model
consisting of galactic and extra-galactic sources \cite{waxman}, that
includes the GZK effect.  We use the extra-galactic source model of
Berezinsky {\it et al.} \cite{kn:berezinsky}, where we assume a uniform
(over the Universe) proton source distribution with a maximum at-source
energy of $10^{21}$ eV, and a galactic spectrum consistent with observations
that the composition changes from heavy to light near $10^{18}$ eV. The
$\chi^2$ of this fit is 48.5 for 37 degrees of freedom, and the fit is shown
in Fig.~\ref{spec1}.  Details can be found in \cite{hires_astroph}.  In this
model the fall-off above $\log{E}$ of 19.8 is due to crossing the pion
production threshold, and the second knee comes from $e^+ e^-$ pair
production pile-up.

For comparison, the published AGASA spectrum~\cite{agasa} is shown in
Fig.~\ref{spec1}. Compared to the HiRes monocular spectrum, the AGASA flux
values are about 60-70\% higher. The AGASA data contain 11 events above
$10^{20}$~eV and 24 above $10^{19.8}$~eV, with an integrated exposure of
$\sim{1.0}\times{10}^{3}$~km$^{2}$sr-yr. The HiRes data have two
events at or above $10^{20}$~eV, and 10 above $10^{19.8}$~eV, with exposures
of $2.4$ and $2.2\times{10}^{3}$~km$^{2}$sr-yr.
at these two energies. If we were to increase the HiRes energy scale by
$17$\% (one RMS systematic deviation), the number of events above
$10^{20}$, and $10^{19.8}$~eV would become three and 20. A decrease in
energy scale of $17$\% changes these numbers to one and 6.

This work is supported by US NSF grants PHY-9321949 PHY-9974537,
PHY-9904048, PHYS-0245428, PHY-0140688, by the DOE grant
FG03-92ER40732, and by the Australian Research Council.  We gratefully
acknowledge the contributions from the technical staffs of our home
institutions and the Utah Center for High Performance Computing. The
cooperation of Colonels E.~Fischer and G.~Harter, the US Army, and the
Dugway Proving Ground staff is greatly appreciated.

\end{document}